\documentstyle[aps,prl,epsfig]{revtex}
\def\Journal#1#2#3#4{{#1} {\bf #2} (#4) #3}

\def\NPB{{\em Nucl. Phys.} B}

\def\PRL{\em Phys. Rev. Lett.}

\def\PRC{{\em Phys. Rev.} C}
\def\PRD{{\em Phys. Rev.} D}

\def\be{\begin{equation}}
\def\ee{\end{equation}}
\def\bea{\begin{eqnarray}}
\def\eea{\end{eqnarray}}

\begin{document}

\title{
Study of $N^*$ Production from $J/\psi \to p\bar{p}\eta $}

\author{
J.~Z.~Bai$^{1}$,    Y.~Ban$^{6}$,      J.~G.~Bian$^{1}$,
J.~F.~Chang$^{1}$,   A.~D.~Chen$^{1}$,
H.~F.~Chen$^{2}$,   H.~S.~Chen$^{1}$,   J.~C.~Chen$^{1}$,  
X.~D.~Chen$^{1}$,
Y.~B.~Chen$^{1}$
% ...
\\(BES Collaboration)\cite{besjpsi}\\
H.C.Chiang,$^{1}$ G.X.Peng,$^{1}$ J.X.Wang,$^{1}$ J.J.Zhu,$^{2}$ }

\vspace{1cm}

\address{
$^{1}$ Institute of High Energy Physics, Beijing 100039,
People's Republic of China \\
$^{2}$ University of Science and Technology of China, Hefei 230026,
People's Republic of China \\
$^{3}$ Shandong University, Jinan 250100,
People's Republic of China \\
$^{4}$ Hangzhou University, Hanzhou 310028,
People's Republic of China \\
$^{5}$ Shanghai Jiaotong University, Shanghai 200030,
People's Republic of China \\
$^{6}$ Peking University, Beijing 100871,
People's Republic of China \\
$^{7}$ Hua Zhong Normal University, Wuhan 430079,
People's Republic of China \\
$^{8}$ China Center for Advanced Science and Technology(CCAST), World
Laboratory, Beijing 100080, People's Republic of China\\
$^{9}$ Nankai University, Tianjin 300071,  
People's Republic of China \\
$^{10}$ Henan Normal University, Xinxiang 453002,
People's Republic of China \\
$^{11}$ Wuhan University, Wuhan 430072,  
People's Republic of China \\
$^{12}$ Hunan University, Changsha 410082,
People's Republic of China \\
$^{13}$ Liaoning University, Shenyang 110036,
People's Republic of China }

\maketitle

\begin{abstract}

Data are presented on the reaction $J/\psi\to p\bar p\eta$ using
$7.8\times 10^6 J/\psi$ triggers collected by the
BEjing Spectrometer (BES). A partial wave analysis is performed.
A clear enhancement near the $p\eta(\bar p\eta)$ threshold is observed.
It is fitted with a $J^P=\frac{1}{2}^-$ resonance with mass
$M = 1530\pm 10$ MeV and width $\Gamma = 95\pm 25$ MeV.
In addition, there is a peak around 1650 MeV with $J^P=\frac{1}{2}^-$
preferred also, fitted with $M = 1647\pm 20$ MeV and 
$\Gamma = 145^{+80}_{-45}$ MeV. 
These two $N^*$ resonances are believed to be the two 
well established states, $S_{11}(1535)$ and $S_{11}(1650)$, respectively.
It is the first partial wave study of the production of these resonances
from $J/\psi$ decays. 

\end{abstract}

\vspace{0.2cm}


\vspace{0.2cm}

\section{Introduction}
Nucleons are the most common form of hadronic matter on the earth
and probably in the whole universe. To understand the internal quark-gluon
structure of nucleon and its excited states $N^*$'s is one of the most
important tasks in nowadays particle and nuclear physics. 
The main source of information for the nucleon internal structure
is their mass spectrum, various production and decay rates.
Our present knowledge on this aspect
came almost entirely from the old generation of $\pi N$ experiments
of more than twenty years ago ~\cite{pdg}.
Considering its importance for the understanding of the nonperturbative
QCD~\cite{isgur},
a new generation of experiments on $N^*$ physics with electromagnetic probes
(real photon and space-like virtual photon) has recently been started at new
facilities such as CEBAF at JLAB, ELSA at Bonn, GRAAL at Grenoble.

\begin{figure}[htbp] 
\centerline{\epsfig{file=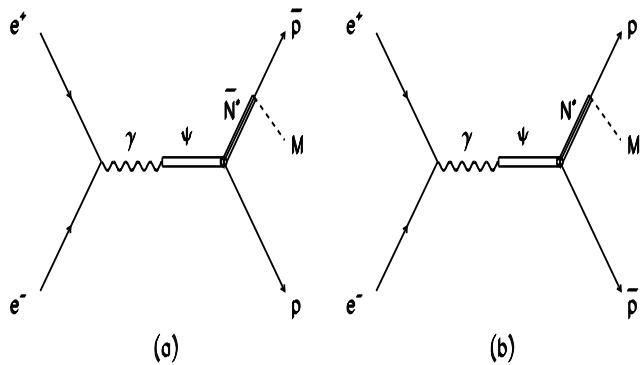,height=3.in,width=4.0in}}
\caption[]{Feynman graphs for $N^*$ and $\bar{N^*}$ production from
$e^+e^-$ collision through $J/\psi$ meson}
\label{figfy}
\end{figure}

The $J/\psi$ experiment at the Beijing Electron-Positron Collider(BEPC)
has long been known as the best place for looking for glueballs. But
in fact it is also an excellent place for studying $N^*$ resonances
~\cite{zou}, especially in the mass range 1-2 GeV.
The corresponding Feynman  graphs for the $N^*$ and $\bar{N^*}$
production are shown in Fig.~\ref{figfy}. These graphs are almost
identical to those describing the $N^*$ electro-production process
if the direction of the time axis is rotated by $90^o$. 
The only difference is that the virtual photon here is time-like 
instead of space-like and couples to $N N^*$ through a real vector 
charmonium meson $J/\psi$. Therefore almost all channels of $N^*$ decays
studied at the CEBAF and other $\gamma p(ep)$ facilities can also be
studied here. 
                
Among many interesting channels, the $J/\psi \to p \bar{p} \eta $
is a relatively simple one to begin with. According to the
information~\cite{pdg,kru,ben,Svarc} from $\pi N \to \eta N$ and
$\gamma N \to \eta N$ experiments, as well as some early quark shell model
calculation~\cite{mit}, only $N^*(1535) S_{11}$ has a big decay branching
ratio to $N\eta$, while $N^*(1650) S_{11}$ may have a branching ratio
to $N\eta$ of up to 10\% and other $N^*$ resonances below 2.0 GeV have
much smaller branching ratios to the $N\eta$.
 
In this paper, we present BES data and a partial wave analysis on the 
$J/\psi\to p \bar{p} \eta$ reaction. The $N^*(1535)$ and $N^*(1650)$ are 
observed.
This is the first partial wave study of the production of these $N^*$
resonances from
the $J/\psi$ hadronic decays in the world. 
The new information on $J/\psi N N^*$ couplings provides a new source for
studying baryon structure\cite{Zou}.  

\section{BES Detector}

The analysis in this paper uses $7.8 \times
10^6 ~J/\psi$ triggers  collected by the Beijing Spectrometer(BES).
BES is a conventional solenoidal magnet detector that is described in detail
in Ref.~\cite{bes}. A four-layer central drift chamber(CDC) surrounding
the beampipe provides trigger information. A forty-layer cylindrical main
drift chamber(MDC), located radially outside the CDC, provides trajectory
and energy loss ($dE/dX$) information for charged tracks over $85\%$ of the
total solid angle. The momentum resolution is $\sigma_P/P = 0.017\sqrt{1 +
P^2}$($p$ in GeV/c), and the $dE/dX$ resolution for hadron tracks is
$~11\%$.
An array of 48 scintillation counters(with inner radius of 1.157 meter and
outer radius 1.207 meter)surrounding the MDC measures the
time-of-flight(TOF) of charged tracks with a resolution of $~450$ ps for
hadrons. Radially outside of TOF system is a 12 radiation length thick,
lead-gas barrel shower counter(BSC) operating in the limited streamer mode.
This device covers ~ 80\% of the total solid angle and measures the energies
of electrons and photons with an energy resolution of $\sigma_E/E =
22\%/\sqrt{E}$ ($E$ GeV). Outside the BSC is a solenoid, which provides
a 0.4 Tesla magnetic field over the tracking volume. An iron flux return
is instrumented with three double layers of counters that identify muons
of momentum greater than 0.5 GeV/C.
\section{Event Selection}

The $\eta$ is detected here in its $\gamma\gamma$ decay mode.
Therefore, much effort has been devoted to the selection of the events in the 
$2\gamma p \bar{p}$ final state. Each candidate event is required to
have two oppositely signed charged tracks with a good helix fit in the polar
angle range $-0.8 < \cos\theta < 0.8$ in MDC and at least 2 reconstructed
$\gamma$'s in BSC. A vertex is required within an interaction region
$\pm 15$ cm longitudinally and 2 cm radially.   
A minimum energy cut of 60~MeV is imposed on the photons. 
Showers associated with charged tracks are also removed.

\begin{figure}[htbp] 
\centerline{\epsfig{file=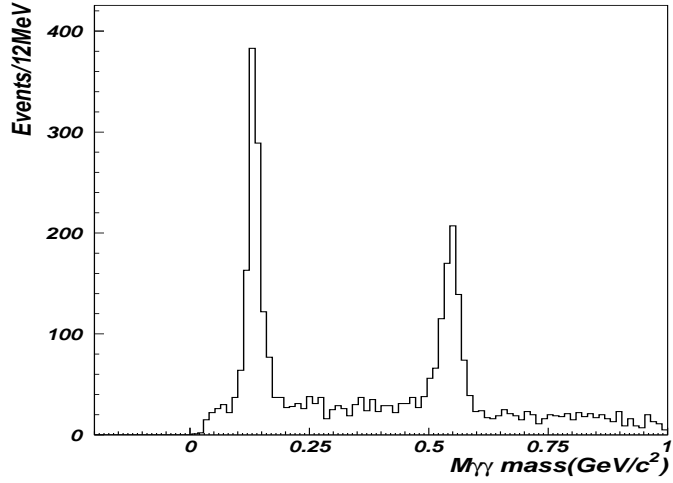,height=3.in,width=4.0in}}
\caption[]{$\gamma \gamma$ invariant mass spectrum after 4C fit for $J/\psi
\rightarrow p\bar{p}\gamma\gamma$}
\label{fige1pe}
\end{figure}

After above selection, we use TOF information to identify the $p\bar{p}$
pairs, and at least one track with unambiguous TOF information is required.
The open angle of two charged tracks smaller than $175^o$ is required in order
to remove back to back events; to remove radiative Bhabha events, we require
$(E_+/P_+ -1)^2+(E_-/P_- -1)^2 >0.3$, where $E_+$, $P_+$ ($E_-$, $P_-$)
are the energy deposited in BSC and momentum of positron (electron)
respectively.      
Events are fitted kinematically to the 4C hypotheses
$J/\psi \to 2\gamma p\bar{p}$. If the number of the
selected photons is larger than two, the fit is repeated using all
permutations of the photons. For events with a good fit, the two 
photon combination with the largest probability is selected.
Fig.~\ref{fige1pe} shows the 
invariant mass spectrum of two gammas, we can see clear $\pi^0$ and
$\eta$ signals.
Meanwhile, the events are also fitted to $J/\psi \to \gamma p\bar{p}$ and 
$4\gamma p \bar{p}$. 
We require 
$$Prob(\chi^2_{(2\gamma  p\bar{p})}, 4C) > Prob(\chi^2_{(\gamma  p\bar{p})}
, 4C),
~~~ Prob(\chi^2_{(2\gamma  p\bar{p})}, 4C) > Prob(\chi^2_{(4\gamma
p\bar{p})},4C)$$ 
to reject the $\gamma  p\bar{p}$ and $p\bar{p}\pi^0\pi^0$ backgrounds.
In order to suppress further the backgrounds with a $\pi^0$, a 5C fit
for the $p\bar p\eta$ with $\eta\to\gamma\gamma$ is
performed on the selected events.
$Prob(\chi^2_{p\bar{p}\eta},5C) > 1\%$  is required. This
5C fit helps improving the mass resolution for combinations of charged 
particles. There are 765 events which survive our selections.

\begin{figure}[htbp] 
\centerline{\epsfig{file=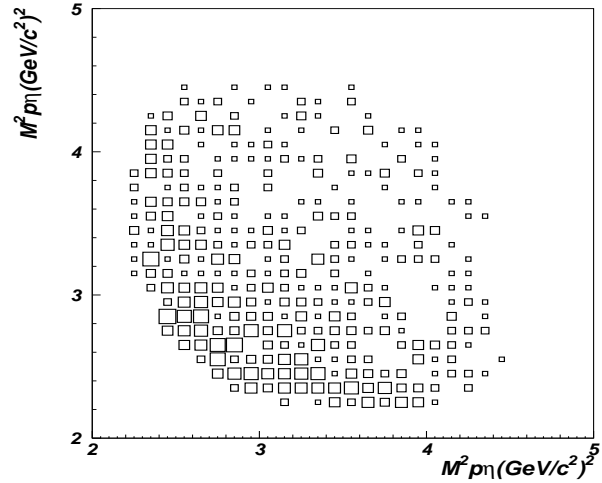,height=3.in,width=3.5in}}
\caption[]{Dalitz plot for $J/\psi \rightarrow p\bar{p}\eta$
($M^2_{p\eta}$ v $M^2_{\bar{p}\eta}$)}           
\label{da2pe}
\end{figure}

\begin{figure}[htbp] 
\centerline{\epsfig{file=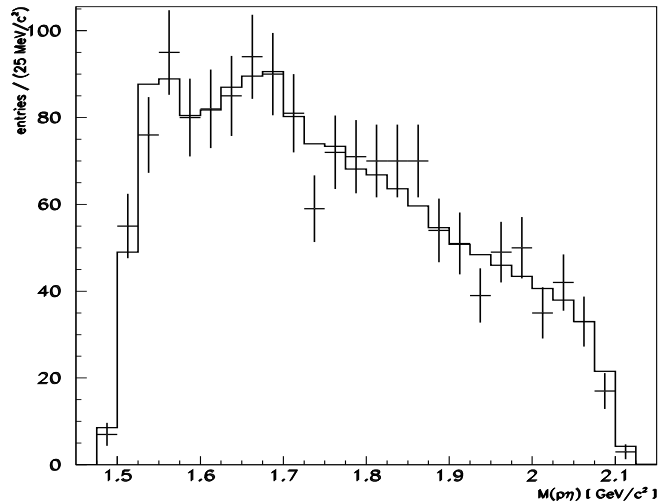,height=3.00in,width=3.8in}}
\caption[]{$p\eta(\bar p\eta)$ invariant mass spectrum for 
$J/\psi\rightarrow p\bar{p}\eta$, crosses are data and histogram the fit}
\label{fig2pe}
\end{figure} 

Fig.~\ref{da2pe} is the Dalitz plot for the decay $J/\psi \to
p\bar{p}\eta$. The corresponding $p\eta(\bar p\eta)$ invariant mass
spectrum is shown in Fig.~\ref{fig2pe}.
There is a clear enhancement near the $p\eta(\bar p\eta)$ threshold.
There are also some bumps around 1.65 GeV and 1.8 GeV. 

\section{Simulation of Signals and Backgrounds}

In order to estimate the selection efficiency, a phase space generator is
used to produce a sample of $p\bar{p}\eta$
events with full detector simulation,  it is 18\% after the Monte Carlo
data go through the same cuts as real data.  

\begin{figure}[htbp] 
\centerline{\epsfig{file=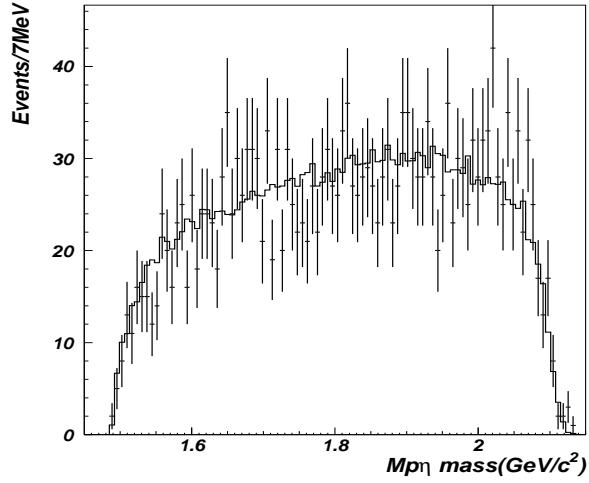,height=3.in,width=3.5in}}
\caption[]{The $p\eta$ invariant mass distribution of the 
background from $J/\psi \to p\bar{p}\pi^0\pi^0$ channel (crosses),
compared with the phase-space distribution for the $J/\psi\to p\bar{p}\eta$ 
reaction (histograms)} 
\label{bg}
\end{figure}

Since
$2\gamma p\bar{p}$ is the final state of the reaction channel, the possible
background may come from $J/\psi \to p\bar{p}\pi^0$, $p\bar{p}\pi^0\pi^0$,
$p\bar{p}\omega$, $\pi^+\pi^-\pi^0$ and $K^+K^- \pi^0$ decay modes.  
For each channel, 10,000 MC events are generated. These MC 
events go through the same analysis program as for the real data. 
We find that only $p\bar{p}\pi^0\pi^0$ channel gives a significant
contribution to the background. The estimated background is 8\%, 

A correct estimation of the background invariant mass shape is essential
to the partial wave analysis. Fortunately, the invariant mass spectrum of
these events is of the same shape as the $J/\psi \to p\bar{p}\eta$
phase space distribution. Fig.~\ref{bg} shows the mass distributions of
the background from $J/\psi \to p\bar{p}\pi^0\pi^0$, compared with the
phase space distribution for the $J/\psi \to p\bar{p}\eta$
process from Monte Carlo data.  A side-band method is also used to
check the shape of possible backgrounds from real data; the same result
is obtained.

\section{Amplitude Analysis}
Based on the study of $p\bar{p}$
and $p\eta$ invariant mass distributions in our data, 
we are mainly interested in the structures at 1535 and 1650 MeV of
the $p\eta$ invariant mass.
According to PDG\cite{pdg}, below 1900 MeV there are eight $N^*$
resonances observed by other experiments, {\sl i.e.}, $P_{11}(1440)$,
$S_{11}(1535)$, $S_{11}(1650)$, $D_{15}(1675)$, $F_{15}(1680)$,
$D_{13}(1700)$, $P_{11}(1710)$ and $P_{13}(1720)$ states, among which
only $S_{11}(1535)$ and $S_{11}(1650)$ were observed in the $\eta N$ decay
mode. Although the two $p\eta$ structures at 1535 and 1650 MeV in our data
are most probably due to $S_{11}(1535)$ and $S_{11}(1650)$ states, we
perform the partial wave analysis by allowing all possible quantum numbers
of $J^P={1\over 2}^\pm$, ${3\over 2}^\pm$ and ${5\over 2}^\pm$.  

The background from multi-$\pi^0$ is $\sim~8\%$ in the 5C fit. 
We have included a phase space background in the PWA fit to allow for
this. 

\begin{figure}[htbp] 
\centerline{\epsfig{file=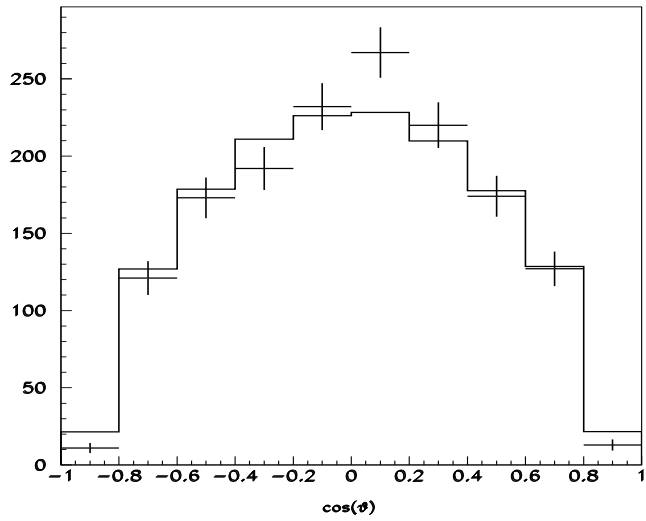,height=3.00in,width=3.8in}}
\caption[]{The polar angle of proton measured with respect to the 
beam direction, crosses are data and histograms the fit}
\label{cos}
\end{figure}

\begin{figure}[htbp] 
\centerline{\epsfig{file=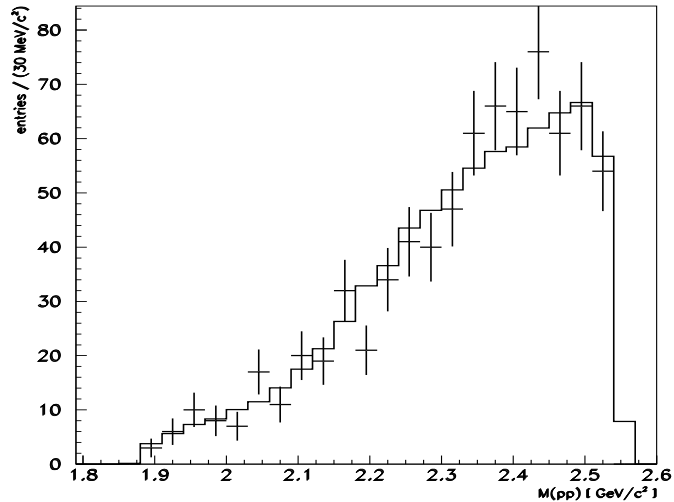,height=3.00in,width=3.8in}}
\caption[]{$p\bar{p}$ invariant mass spectra for $J/\psi \rightarrow
p\bar{p} \eta$, crosses are data and histograms the fit}
\label{ppbar}
\end{figure}

In this analysis, we use the effective Lagrangian
approach for the partial wave analysis. The relevant spin-1/2 interaction
Lagrangians are\cite{Nimai,Olsson}
\bea
{\cal L}_{\eta PR}^{1} &=& -ig_{\eta PR}\bar P\Gamma R\eta + H.c. , \\
{\cal L}^{(1)}_{\psi PR} &=& \frac{ig_{T_R}}{M_R+M_P}\bar R\Gamma_{\mu\nu}
q^\nu P\psi^\mu +H.c. , \\
{\cal L}^{(2)}_{\psi PR} &=& -g_{V_R}\bar R\Gamma_\mu P\psi^\mu +H.c.
\eea
where $R$ is the generic notation for the resonance with mass $M_R$, $P$
for proton with mass $M_P$ and $\psi$ for $J/\psi$ with four-momentum $q$.
The vertex coupling constants $g_{\eta PR}$, $g_{T_R}$ and $g_{V_R}$ are
parameters to be determined by fitting experimental data.
Note here from the single $\bar pp\eta$ channel, we can only determine
products of coupling constants, {\sl i.e.}, $g_{\eta PR}g_{T_R}$ and 
$g_{\eta PR}g_{V_R}$.
The operator structures for the $\Gamma$, $\Gamma_\mu$ and
$\Gamma_{\mu\nu}$ are
\be
\Gamma=1, \quad \Gamma_\mu=\gamma_5\gamma_\mu, \quad
\Gamma_{\mu\nu}=\gamma_5\sigma_{\mu\nu},
\label{eq:4}
\ee
\be
\Gamma=\gamma_5, \quad \Gamma_\mu=\gamma_\mu, \quad
\Gamma_{\mu\nu}=\sigma_{\mu\nu},
\label{eq:5}
\ee
where (\ref{eq:4}) and (\ref{eq:5}) correspond to nucleon resonances of
$J^P={1\over 2}^-$ and ${1\over 2}^+$, respectively.
The interaction Lagrangians for $J^P={3\over 2}^\pm$ and 
${5\over 2}^\pm$ $N^*$ resonances are constructed similarly\cite{Rarita}.
The amplitudes in the PWA analysis are constructed from
these Lorentz-invariant interactions and the $N^*$
propagators\cite{Fronsdal} for $J/\psi$ initial states with helicity 
$\pm 1$.  The relative magnitudes and phases of the 
amplitudes are determined by a maximum likelihood fit to the data. 
The BW parameters for different states
are fitted. The $p\eta$ mass projection fitted
to the real data is shown in Fig.~\ref{fig2pe}.  Fig.~\ref{cos} shows the
data and fit for the polar angle of the proton measured with respect to
the beam direction.
Fig.~\ref{ppbar} shows the invariant mass spectra for $p\bar{p}$.
We now discuss the features of the data and the outcome of fits.

\vspace{0.3cm}

\underline{$S_{11}(1535)$}\\
The peak around 1535 MeV near the $p\eta$ threshold optimizes at
$M=1530\pm 10$ MeV and $\Gamma = 95\pm 15$ MeV.
The data favour $J^{P} = \frac{1}{2}^{-}$ over others.
A fit with $J^{P} =  \frac{1}{2}^{+}$ instead gives $\ln~L$ worse by 7.5 
than for $\frac{1}{2}^{-}$ assignment (Our definition of $ln~L$ is such
that it increases by 0.5 for a one standard deviation change in one 
parameter). Fits with other quantum numbers are much worse, {\sl e.g.},
$\ln L$ worse by 26 for $J^{P} =  \frac{3}{2}^{-}$.  
With our 4 fitted parameters, the statistical significance
of the peak is above 6.0$\sigma$.  
It is obviously the $S11$ $N^*(1535)$ resonance. It makes the largest
contribution $(56\pm 15)\%$ to the $p\bar p\eta$ final states, the errors
here and later include both statistics and systematic errors from the fit.
Our results for $N^*(1535)$ are consistent with the resonance parameters
of PDG\cite{pdg} and a recent detailed analysis of $\pi N$ $S_{11}$
partial wave by Vrana, Dytman and Lee\cite{Dytman}.

\underline{$S_{11}(1650)$}\\
The second peak around 1650 MeV is also fitted with a $J^{P}=\frac{1}{2}^-$
resonance $N^*(1650)$. It optimizes at $M=1647\pm 20$ MeV,
$\Gamma = 145^{+80}_{-45}$ MeV, and
contributes $(24^{+5}_{-15})\%$ to the $p\bar p\eta$ final states. 
We have tried fits to this peak with other quantum numbers.
The log likelihood is worse by 3, 4, 5, 6, 12 for ${5\over 2}^+$,
${3\over 2}^+$, ${5\over 2}^-$, ${1\over 2}^+$ and ${3\over 2}^-$,
respectively. Note there are two more free parameters for 
$J^P={3\over 2}^\pm$ and $J^P={5\over 2}^\pm$ modes than for 
$J^P={1\over 2}^\pm$ modes.
If we assume no contribution from resonance around 1650 MeV, then the mass
and width of $S_{11}(1535)$ optimize around 1570 MeV and 270 MeV,
respectively, which are 
out of the range of PDG values. The log likelihood is worse by 14. 

A small improvement to the fit is given by including a $J^{P}=\frac{1}{2}^+$
resonance, which optimizes at $M=1800\pm 40$ MeV and 
$\Gamma = 165^{+165}_{-85}$ MeV. It contributes $(12\pm 7)\%$ to the
$p\bar p\eta$ final states.
The statistical significance of the peak is only $2.0 \sigma$. We have
tried other quantum numbers. All of them give equally good fits within one
standard deviation.
There is a theoretical prediction\cite{Page} that the lowest-lying 
hybrid states should be around this energy region.

We also include a small contribution $(8\pm 4)\%$ from the tail of
$P_{11}(1440)$ resonance with its mass and width fixed to its PDG central
values for its pole position\cite{pdg}. It is also only a $2.0 \sigma$ 
effect.

\begin{figure}[htbp] 
  \begin{minipage}[b]{0.415\linewidth}
\centerline{\epsfig{file=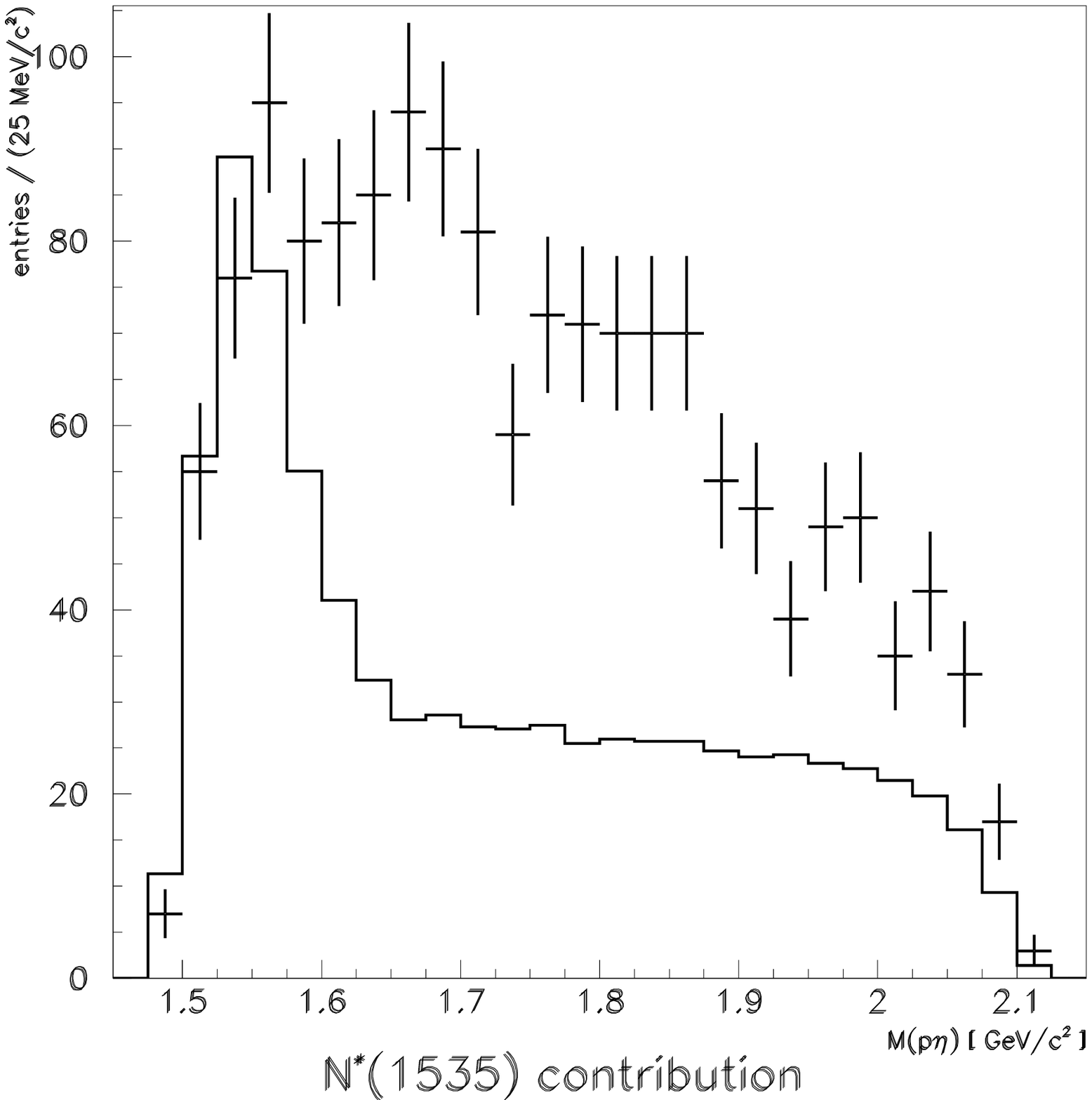,height=2.450in,width=3.in}}
 \end{minipage}
\hfill
  \begin{minipage}[b]{0.415\linewidth}
\centerline{\epsfig{file=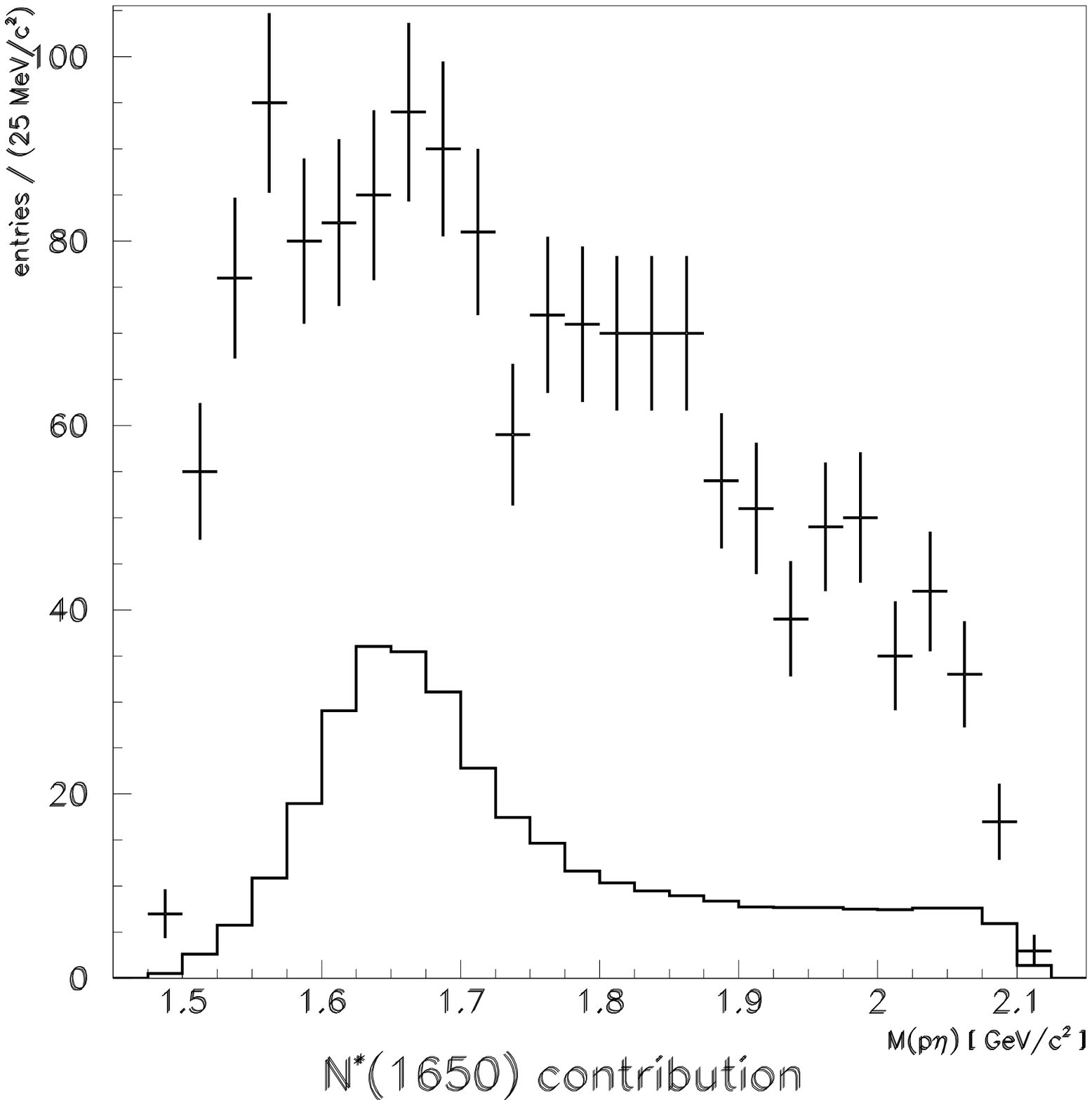,height=2.450in,width=3.in}}
 \end{minipage}
\hfill
  \begin{minipage}[b]{0.415\linewidth}
\centerline{\epsfig{file=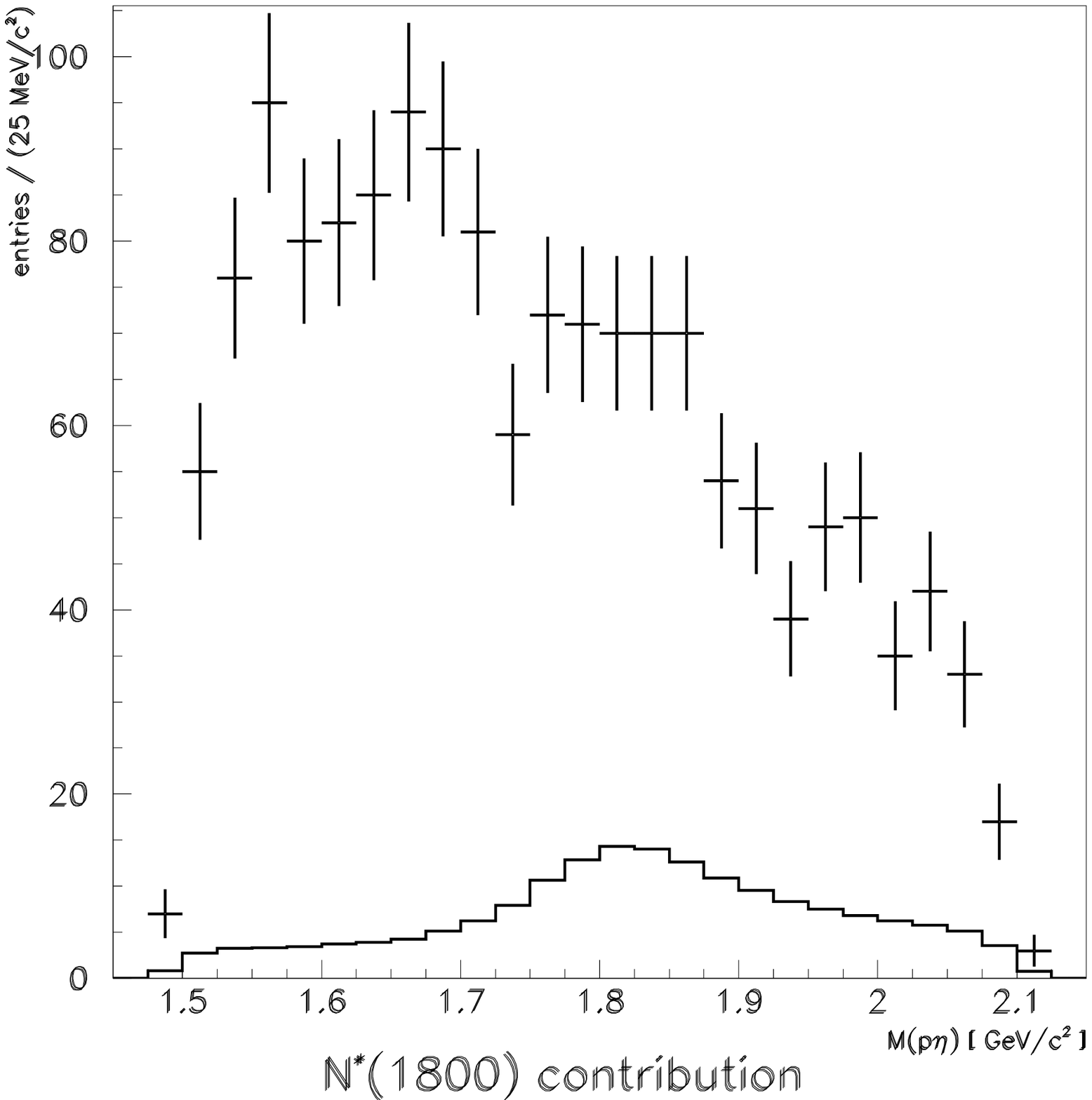,height=2.450in,width=3.in}}
 \end{minipage}
\hfill
  \begin{minipage}[b]{0.415\linewidth}
\centerline{\epsfig{file=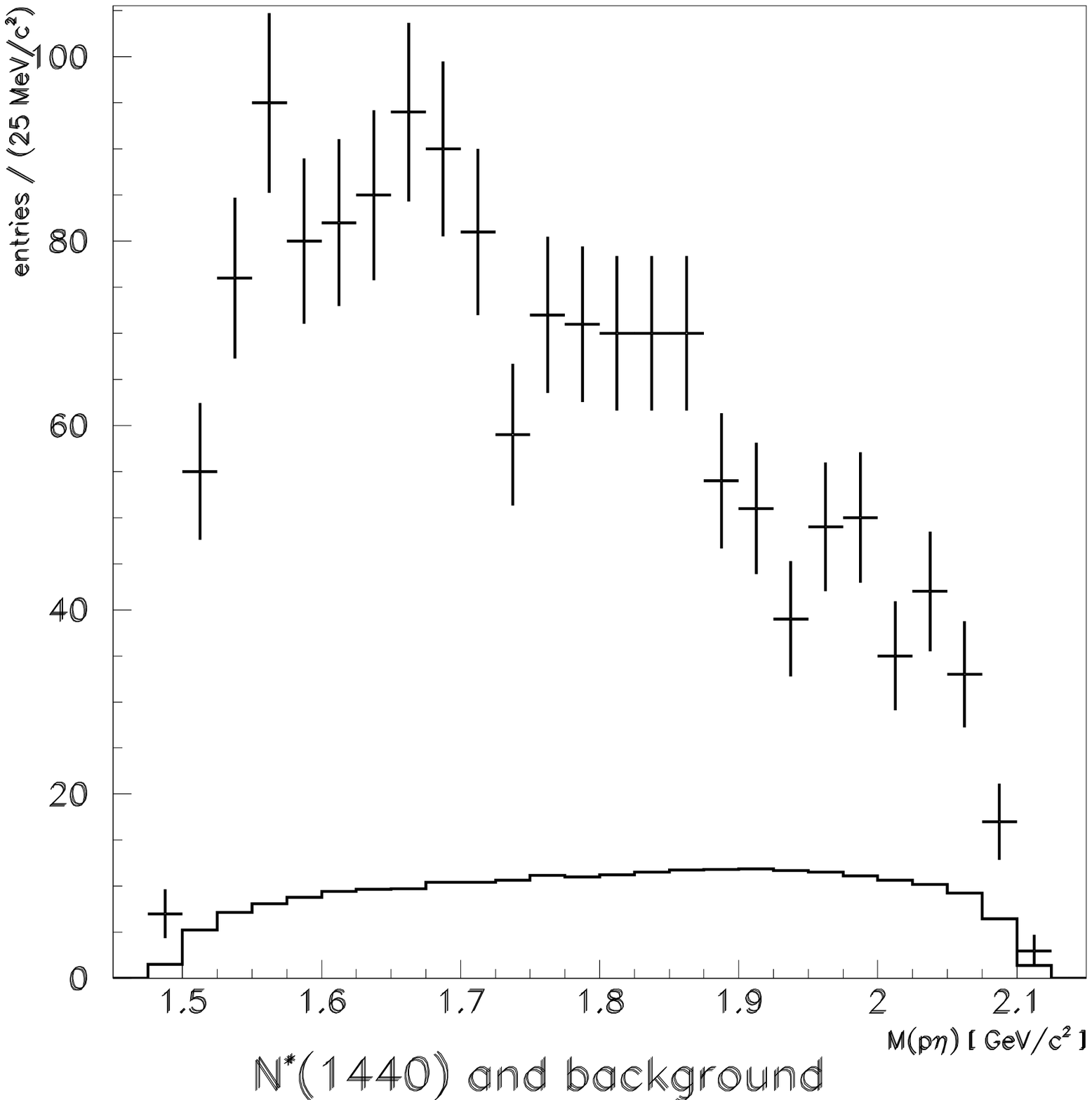,height=2.450in,width=3.in}}
 \end{minipage}
\caption[]{Component contribution }
\label{pro}
\end{figure}

Fig.~\ref{pro} shows the different component contribution.



An interesting result is that the ${\cal L}^{(2)}_{\psi PR}$ term given
by Eq.(3) makes insignificant contribution for both $S_{11}(1535)$
and $S_{11}(1650)$. If we drop this kind of couplings for both resonances,
the likelihood value for the fit is only worse by 0.9 for 4 less
free parameters. This kind of couplings should vanish for the real
photon coupling to $NN^*$ due to the requirement of the gauge invariance.
Why it also vanishes for the $\psi NN^*$
coupling needs to be understood. A theoretical calculation\cite{Okubo}
assuming pure ${\cal L}^{(2)}_{\psi PR}$ coupling without ${\cal
L}^{(1)}_{\psi PR}$ coupling failed to reproduce the basic feature
of the $J/\psi\to\bar pp\eta$ data. This is consistent with our
observation that the ${\cal L}^{(1)}_{\psi PR}$ coupling dominates
for both $N^*(1535)$ and $N^*(1650)$.


In the Vector Meson Dominance (VMD) picture, the virtual photon couples
to the $NN^*$ through vector mesons, and the electro-magnetic $NN^*$
transition form factors $g_{\gamma^*NN^*}$ can be expressed in terms of
photon-meson coupling strengths $C_{\gamma V}$ and meson-$NN^*$ vertex
form factors $g_{_{VNN^*}}$:
\be
g_{\gamma^*NN^*}(q^2)=\sum_j\frac{m_j^2C_{\gamma
V_j}}{m^2_j-q^2-im_j\Gamma_j}g_{_{V_jNN^*}}(q^2)
\ee
with
\be
C_{\gamma V}=\sqrt{\frac{3\Gamma_{V\to e^+e^-}}{\alpha m_V}}.
\ee
At $q^2=M^2_\psi$, the $J/\psi$ meson dominates. The terms from other
vector mesons are negligible.
From our PWA results here and other relevant information from
PDG\cite{pdg}, we can deduce the transition form factor for the time-like
virtual photon to $PN^*(1535)$ as
\be
|g_{\gamma^*pN^*}(q^2=M_\psi^2)|= 2.3\pm 0.5 ,
\ee
which is related to the more familiar helicity amplitude $A^P_{1/2}$ for
$N^*\to\gamma P$ by
\be
|A^P_{1/2}|^2=\left(\frac{g_{\gamma pN^*}(q^2=0)}{M_{N^*}+M_P}\right)^2
\frac{(M^2_{N^*}-M^2_P)}{2M_P} .
\ee
This is the first measurement of the form factor $g_{\gamma^*pN^*}$
for a time-like virtual photon. It provides a new challenge for various
theoretical models to reproduce it, complementary to the information
for a real or space-like virtual photon\cite{Stoler}.

\section{Conclusion}

In summary,  the $J/\psi$ decay at BEPC provides a new excellent
laboratory for studying the $N^*$ resonances.
We have studied the $J/\psi \rightarrow p \bar{p} \eta$ decay
channel, and a PWA analysis is performed on the data. 
There is a definite requirement for a $J^{P}=\frac{1}{2}^-$ component at 
$M = 1530\pm 10$ MeV with $\Gamma =95\pm 25$ MeV near the $\eta N$
threshold. In addition, there is an obvious resonance around 1650 MeV 
with $J^P=\frac{1}{2}^-$ preferred, $M = 1647\pm 20$ MeV and 
$\Gamma = 145^{+80}_{-45}$ MeV.
In the higher $p\eta$($\bar{p}\eta$) mass
region, there is an evidence for a structure around 1800 MeV;
with present statistics we cannot determine its quantum numbers.

All above analysis is the first step for us to probe $N^*$ baryons at BES.
We will perform a detail study of $N^*$ baryons in the following $J/\psi$
decay channels: $J/\psi \to p \bar{p} \pi^0$, $p\bar n\pi^-$, $p \bar{p}
\pi^0 \pi^0$, $p\bar{p}\pi^+ \pi^-$, $p \bar{p}\eta'$, $p \bar{p}\omega$
and so on.  

With the forthcoming 50 million more $J/\psi$ events in near future, more
precise partial wave analyses can be carried out on many channels
involving $N^*$ resonances and should offer some determinations of
$N^*$ properties. A systematic experimental
study of the $N^*$ production from the $J/\psi$ decays
is underway and will provide a new domain to explore the internal
quark-gluon structure of these excited nucleons.

\section{Acknowledgements}

We thank D.V.Bugg, S.Dytman, L.Kisslinger, P.R.Page and P.Stoler
for useful discussions, and the staff of IHEP for technical support in
running the experiment. This work is partly supported by China
Postdoctoral Science Foundation and National Science Foundation of China
under contract Nos. 19290401, 19605007, 19991487 and 19905011; and by the
Chinese Academy of Sciences under contract No. KJ 95T-03(IHEP)

\end{document}